\documentclass[12pt]{iopart}

\usepackage{graphicx}

\begin{document}

\title[Is a description deeper than the quantum one possible?]{Is a description deeper than the quantum one possible?}

\author{GianCarlo Ghirardi$^1$ and Raffaele Romano$^2$}

\address{$^1$ Department of Physics, University of Trieste, the Abdus Salam ICTP, Trieste, Italy}
\ead{ghirardi@ictp.it}
\address{$^2$ Department of Mathematics, Iowa State University, Ames IA, USA}
\ead{rromano@iastate.edu}

\begin{abstract}

\noindent Recently, it has been argued that quantum mechanics is a complete theory, and that
different quantum states do necessarily correspond to different elements of reality, under the assumptions
that quantum mechanics is correct and that measurement settings can be freely chosen. In this work, we prove
that this result is a consequence of an unnecessarily strong mathematical expression of the free choice
assumption, which embodies more conditions than explicitly stated. The issues of the completeness of
quantum mechanics, and of the interpretation of the state vector, are by no means resolved. Taking this perspective,
we describe how the recently introduced class of crypto-nonlocal hidden variables theories can be used to
characterize the maximal possible departure from quantum mechanics, when the system consists of a pair of qubits.

\end{abstract}


%\keywords{Hidden variables, free will}

\maketitle

%%%%%%%%%%%%%%%%%%%%%%%%%%%%%%%%%%%%%%%%%%%%%%%%%%%%%%%%%%%%%%%%%%%%%%%%

\section{Introduction}

Despite the unprecedented success of quantum theory (and its field-theoretical relativistic
generalization) in explaining any experimental evidence of the microscopic world, the
interpretation of the formalism represents a long-standing problem. This is partially
due to the counter-intuitive features of the description of the micro-world, when
compared to the concepts derived from classical physics, the prominent examples being given
by probabilism, indeterminism and non-locality. Nonetheless, the really unpleasant
feature of the quantum formalism is the fact that it appears more like a set of operational
prescriptions to fit the experimental data, rather than a coherent description of  reality.
The theory relies on two different kinds of evolution depending on the rather vague notion of
measurement, is not able to account for the behavior of the classical world in the limit of macroscopic objects, and,
consequently, its range of validity is not well defined.

The famous incompleteness argument by Einstein, Podolski and Rosen has raised the questions of
whether quantum mechanics is a complete theory or not, and how to interpret the state
description required by the theory in terms of the quantum state vector $\psi$. In
particular, it is unclear whether $\psi$ represents a state of reality or rather a state of
knowledge, as suggested by its updating following a measurement procedure. According to these
lines, several {\it ontological models} of quantum mechanics have been introduced, that is, theories
which are predictively equivalent to quantum mechanics, but providing a possibly richer description of
the microscopic reality through the so-called {\it ontic state}, the most accurate specification of the
physical state of the system, at least in principle. In these theories, the state vector $\psi$ might
embody only partial information on the ontic state, since it is associated to a distribution
$\rho_{\psi} (\lambda)$ on the space of the ontic variable $\lambda$, with $\rho_{\psi} (\lambda) \geq 0$ and
\begin{equation}\label{norm}
    \int \rho_{\psi} (\lambda) d \lambda = 1 \qquad {\rm for \, all} \, \psi.
\end{equation}
The ontic state $\lambda$ accounts for  the elements of reality of the underlying theory since it provides
the complete description of the state of the system, but, in principle, it might be not fully accessible.
This is the reason why these models were formerly characterized as {\it hidden variables theories}.

Notice that the ontological status of $\psi$ can be understood by studying the distribution $\rho_{\psi} (\lambda)$,
which, in the so-called {\it $\psi$-epistemic models}, has overlapping supports,
while in the so-called {\it $\psi$-ontic models}, it has disjoint supports~\cite{harrigan} for different $\psi$.
Accordingly, in a $\psi$-ontic theory different state vectors necessarily correspond to different ontic
states, whereas in a $\psi$-epistemic theory they could correspond to the same ontic state.
One could associate to $\psi$  well defined elements of reality only in the first case.

In a recent work it has been argued that, if quantum mechanics is correct and the experimenters can
freely choose their own settings, no theory can outperform its predictive power,
that is, the microscopic world can only be described in terms of probabilistic laws, and the probabilities
are definitely those provided by the quantum formalism~\cite{colbeck2}. In other words, quantum mechanics really is
complete. Moreover, as a corollary of this result, it has been proven that $\psi$-epistemic models necessarily
contrast with quantum mechanics~\cite{colbeck}. Accordingly, it has been concluded that $\psi$ does not represents a
state of knowledge but rather a state of reality. This result  already appeared in the recent literature~\cite{pusey},
although limited by the assumption that factorized quantum states correspond to factorized states of the underlying
theory, as highlighted by an explicit model~\cite{lewis}.

In our opinion, the general scenario envisaged in~\cite{colbeck2,colbeck} is   not the right one,
since the mathematical expression of the free choice assumption, denoted by $FR$, is unnecessarily strong.
While a general criticism to these works from a more epistemological perspective has been presented in
Ref.~\cite{ghirardi}, here we prove that $FR$ embodies more than the free choice assumption.
Therefore, we provide evidence that the argument put forward in~\cite{colbeck2,colbeck} is not conclusive for
stating the completeness of quantum mechanics, nor in determining the ontological status of the quantum state
vector.

This result triggers the following question: are there theories which are compatible with quantum mechanics,
but potentially distinguishable from it? We argue that this question has a positive answer, and the theories
fulfilling this requirement are exactly the recently introduced crypto-nonlocal hidden variables models.
We discuss what is the maximal departure from quantum expectations that these models can provide in the simple
case of a pair of two-level systems.

\section{Free will and the ontological status of quantum mechanics}

As already anticipated, Ref.~\cite{colbeck2} derives completeness of quantum mechanics and the one-to-one correspondence between quantum state
vectors and elements of reality  from the assumptions $QM$ of the validity of the predictions of quantum mechanics, and a
request $FR$ which, according to the authors,  expresses the freedom of choosing the measurement settings.  One considers two space-like separated observers performing
local measurements on the two parties of an entangled state $\psi$. The
measurement settings are given by vectors $A$ and $B$, the outcomes are denoted
by $X$ and $Y$. Following~\cite{colbeck2,colbeck}, we assume that additional information on the ontic
state  $\lambda$ is available and can be obtained through a measurement with setting $C$ and output $Z$.
In the following, we do not exclude the case $Z = \lambda$, which means that the ontic state is fully accessible.
We consider all these quantities as random variables. The $FR$ assumption is the condition that
\begin{quote}
{\it [...] the input $A$ can be chosen to be uncorrelated with all
the space-time random variables whose coordinates lie outside the future light-cone of its coordinates}~\cite{colbeck2},
\end{quote}
and the same requirement holds also for $B$ and $C$. The authors of~\cite{colbeck2} have
expressed this assumption by imposing the following constraints on the
conditional probabilities:
\begin{equation}
P_{A|BCYZ} = P_A, \quad
P_{B|ACXZ} = P_B, \quad
P_{C|ABXY} = P_C,
\end{equation}
which are all needed to derive the main results of~\cite{colbeck2,colbeck}.
However, we notice that the free will condition is consistent, among others, with a condition weaker than $FR$,
which makes reference exclusively to the fact that the two observers can independently choose
which observables to measure:
\begin{equation}
P_{A|B\lambda} = P_A, \quad P_{B|A\lambda} = P_B,
\end{equation}
where $\lambda$ is the aforementioned ontic state. This condition, denoted by $FW$
in the following, produces the relevant factorization $P_{AB\lambda} = P_A P_B P_{\lambda}$.
Meaningfully, $FW$ is unrelated with the physically important assumption that the two observers
cannot communicate superluminally, denoted as $NS$, and expressed by
\begin{equation}
P_{X|AB} = P_{X|A}, \quad P_{Y|AB} = P_{Y|B}.
\end{equation}
This is reasonable: one could imagine artificial models in which free will and superluminal signalling coexist.
Notice that $FR \Rightarrow NS$, supporting the idea that $FR$ embodies more than the free choice assumption.

We observe that in~\cite{colbeck2} it is pointed out that the information supplementing $\psi$
\begin{quote}
{\it [...] must be static, that is, its behavior cannot depend on
where or when it is observed}.
\end{quote}
Otherwise said, the region of events corresponding to the acquisition of this
information can be chosen to be space-like with respect to the events associated to $A$ and $B$.
In~\cite{colbeck2}, this statement is presented as a simple remark and it does not constitute a new assumption;
nonetheless, here we choose to denote it by $ST$, and to express it as $P_{CZ|ABXY} = P_{CZ}$. It turns out that
\begin{equation}\label{imply}
    FW \wedge NS \wedge ST \Rightarrow FR.
\end{equation}
In fact, from $ST$ it follows that $P_{ABY|CZ} = P_{ABY}$;
moreover we have
\begin{equation}\label{eq1}
    P_{ABY|CZ} = P_{A|BYCZ} P_{BY|CZ}
                           = P_{A|BYCZ} P_{BY}
\end{equation}
by using again $ST$, but also
\begin{equation}\label{eq2}
    P_{ABY} = P_{AB} P_{Y|AB} =  P_A P_B P_{Y|B}
                           = P_A P_{BY}
\end{equation}
from $NS$ and $FW$. By comparing (\ref{eq1})
and (\ref{eq2}) we find that $P_{A|BYCZ} = P_A$, and a similar argument proves that
$P_{B|AXCZ} = P_B$. Finally, $P_{C|AXBY} = P_C$ is a direct implication of $ST$.
This result clearly shows that $FR$ corresponds to more than the
free will of the observers. Notice that, if we assume $Z = \lambda$, we can
prove also the converse of implication (\ref{imply}), meaning that, if the ontic
state would be completely accessible, $FR$ would be  equivalent to the conjunction of the conditions
$FW$, $NS$ and $ST$. For a detailed account of the role of the accessibility of
$\lambda$ in this analysis, see~\cite{ghirardi}.

Therefore, violation of $FR$ does not necessarily imply lack of free will
as long as $ST$ or $NS$ are violated. This means that ontological models fully consistent
with quantum mechanics, with the free will assumption (expressed through $FW$) and without superluminal
communication are indeed possible, as long as the supplementary information on the ontic state is not static.
Moreover, these models could be made of the $\psi$-epistemic type, for instance, by following the lines
described in~\cite{lewis}.

We want to comment about our condition $FW$. We do not consider it as the ultimate expression of the free will assumption,
but only a meaningful substitute to $FR$, which enables us to raise our criticism to the $FR$ assumption. As $FR$, also $FW$
relies on conditional probabilities, and, in our opinion, this is not the most appropriate way to express the free will.
Our target here is only to prove that the conclusions of~\cite{colbeck2,colbeck} are not appropriate, rather than providing
an accurate mathematical expression of the free will assumption.
Notice that the general approach to free will has been expressed by J.S. Bell as
\begin{quote}
{\it for me this means that the values of such variables have implications
only in their future light cones}~\cite{bell},
\end{quote}
and, in our opinion, neither $FR$ nor $FW$ are able to properly express this fact,
since lack of correlations is stronger than lack of implications.

\section{Beyond quantum mechanics: the role of crypto-nonlocal hidden variables models}

Following our reasoning, we conclude that the issue about completeness of quantum mechanics is
still open, and similarly there are not conclusive conclusions concerning the ontological status of the vector
$\psi$, which represents the state of the system in quantum mechanics. We now focus on the following question:
could there be a theory, predictively equivalent to quantum mechanics, but experimentally distinguishable from it?
In other words, would it be possible that a more refined knowledge of $\lambda$ could produce different outcomes (e.g.
different statistics) from quantum mechanics, consistently with the fact that, if the information on the state reduces to that encoded
in $\psi$, these outcomes are exactly those of quantum mechanics? Of course, one has to further
constrain the theory in order to avoid physical inconsistencies. More explicitly, any information on
$\lambda$ cannot be used to implement superluminal communication between distant parties.

It turns out that this requirement is exactly addressed by the class of crypto-nonlocal
hidden variable models, recently introduced by Leggett in a different context (investigation
of non-locality and entanglement of correlated photons)~\cite{leggett}. In the simplified case where the
ontic state, jointly with the settings $a$ and $b$ (the actual values of the random variables $A$ and $B$) determine
the outcomes $x$ and $y$ (the actual values of the random variables $X$ and $Y$)~\footnote{In general, only the
probabilities of these outcomes are determined.}, these models can be described as follows. We express $\lambda$ through two variables $(\mu, \tau)$,
$\mu$ denoting the unaccessible part of the ontic state, and $\tau$ the accessible one.
Now, we can write $\rho_{\psi} (\lambda) = \rho_{\psi, \tau} (\mu)
\rho_{\psi} (\tau)$, and impose that knowledge of $\tau$ does not allow superluminal communication, that is
\begin{eqnarray}\label{locavcnlhv}
    \int x (a, b, \lambda) \rho_{\psi, \tau}(\mu) d \mu &=& f_{\psi}(a, \tau), \nonumber \\
    \int y (a, b, \lambda) \rho_{\psi, \tau}(\mu) d \mu &=& g_{\psi}(b, \tau),
\end{eqnarray}
which are the so-called {\it non-signalling conditions}. The quantities $f_{\psi}(a, \tau)$ and $g_{\psi}(b , \tau)$
are the local averages of the theory at the intermediate level, that is, when the state of the system
is described by $\tau$. Non-locality, which is apparent by the functional dependence $x(a,b,\lambda)$ and $y(a,b,\lambda)$,
has been canceled out. As required, when we additionally average over $\tau$ we recover the quantum expectations,
\begin{eqnarray}\label{locavcnlhv2}
    \int f_{\psi}(a, \tau) \rho_{\psi}(\tau) d \tau &=& \langle x(a) \rangle_{\psi}, \nonumber \\
    \int g_{\psi}(b, \tau) \rho_{\psi}(\tau) d \tau &=& \langle y(b) \rangle_{\psi}.
\end{eqnarray}
The theory is experimentally distinguishable from quantum mechanics as long as
$f_{\psi}(a, \tau) \ne  \langle x(a) \rangle_{\psi}$ and/or $g_{\psi}(b, \tau) \ne  \langle y(b) \rangle_{\psi}$.

An example of this scheme has been recently described in~\cite{ghirardi2} for a pair of two qubits.
It is a generalization of the famous Bell's model for the singlet state of a pair of two-level systems, valid for an
arbitrary state $\psi$ written as
\begin{equation}\label{genstate}
    \vert \psi \rangle = \sin{\frac{\theta}{2}} \vert 00 \rangle + \cos{\frac{\theta}{2}} \vert 11 \rangle,
\end{equation}
with $\theta \in [0, \pi/2]$. If $\theta = 0$, $\vert \psi \rangle$ is a separable state state;
if $\theta = \pi/2$ it is a maximally entangled state. The ontic state is given by the pair
$\lambda = (\psi, \tilde{\lambda})$, where $\tilde{\lambda}$ is a unit vector in the 3-dimensional
real space. By construction, the model is $\psi$-ontic, because different vectors $\psi$ are
necessarily associated to different ontic states $\lambda$~\footnote{Nonetheless, a $\psi$-epistemic model
can be obtained by suitably modifying this scheme.}. For specific values of the local settings,
$A = a$ and $B = b$ ($a$ and $b$ are real, unit vectors), the local observables are given by $\sigma
\cdot a$ and $\sigma \cdot b$, where $\sigma = (\sigma_x, \sigma_y,\sigma_z)$ is the vector of Pauli
matrices. In particular, $\sigma_z$ is defined so that $\vert 0 \rangle$ and $\vert 1 \rangle$ are
its $+1$ and $-1$ eigenvectors respectively. Without loss of generality we assume that $a$ and $b$
lie in the plane orthogonal to the direction of propagation of the entangled particles, assumed to
depart from a common source. The local measurement outcomes are given by
\begin{equation}\label{defouta}
    x (a, b, \tilde{\lambda}) = \left\{
                                    \begin{array}{ll}
                                      +1, & \hbox{\rm if $\hat{a} \cdot \tilde{\lambda} \geq\cos{\xi}$,} \\
                                      -1, & \hbox{\rm if $\hat{a} \cdot \tilde{\lambda} < \cos{\xi}$,}
                                    \end{array}
                                  \right.
\end{equation}
and
\begin{equation}\label{defoutb}
    y (b, \tilde{\lambda}) = \left\{
                                    \begin{array}{ll}
                                      +1, & \hbox{\rm if $b \cdot \tilde{\lambda} \geq \cos{\chi}$,} \\
                                      -1, & \hbox{\rm if $b \cdot \tilde{\lambda} < \cos{\chi}$.}
                                    \end{array}
                                  \right.
\end{equation}
In the previous relations, $\hat{a} = \hat{a} (a, b)$ is in the plane of $a$ and $b$, as detailed in~\cite{ghirardi2}; moreover,
$\cos{\xi} = - \langle x(a) \rangle_{\psi}$, and $\cos{\chi} = - \langle y(b) \rangle_{\psi}$.

With the additional assumption that $\tilde{\lambda}$ is uniformly distributed on the unit sphere, it is possible
to prove that this model is predictively equivalent to quantum mechanics (see~\cite{ghirardi2} for the details).
Moreover, it belongs to the crypto-nonlocal family. We identify $(\mu, \tau)$ with the spherical coordinates of
$\tilde{\lambda}$: $\mu$ is the azimuthal angle and $\tau$ the polar angle, and the north pole is identified by
the direction of the incoming particle. It is easy to prove that $\rho_{\psi, \tau} (\mu) = 1 / 2 \pi$, and,
by construction, integration over $\mu$ cancels non-locality in local averages. We find that~\cite{ghirardi2}
\begin{equation}\label{result}
  f_{\psi} (a, \tau) = \frac{1}{\pi} \, \cos^{-1}{\left(\frac{2 \langle x(a) \rangle_{\psi}^2}{\sin^2{\tau}} - 1 \right)} - 1,
\end{equation}
if $\vert \tau - \frac{\pi}{2} \vert \leq \xi$ and $f_{\psi} (a, \tau) = - 1$ otherwise,
and a similar relation (with $\xi$ replaced by $\chi$) for $g_{\psi} (b, \tau)$. In general, $f_{\psi} (a, \tau) \ne
\langle x(a) \rangle_{\psi}$ and $g_{\psi} (b, \tau) \ne \langle y(b) \rangle_{\psi}$. Therefore, despite the
model is absolutely artificial, it provides evidence that models compatible with quantum mechanics, but in
principle distinguishable from it, are indeed possible, without violating the free will assumption.

\begin{figure}[t]
  \centering
  \includegraphics[width=8cm]{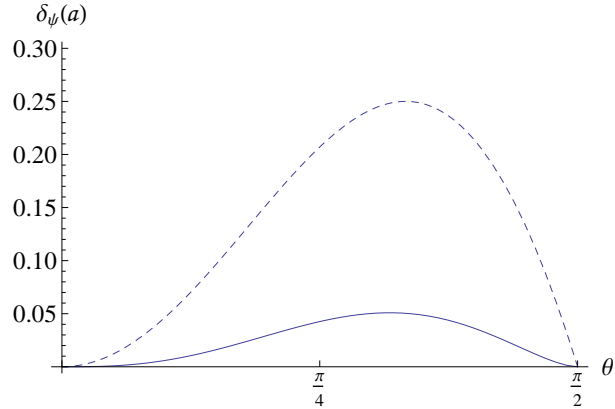}
  \caption{Behavior of $\delta_{\psi} (a)$ as a function of $\theta$. Solid line: $\delta_{\psi} (a)$ for the generalization of the Bell's model
  described in the text; dashed line: upper bound for $\delta_{\psi} (a)$. In both cases we consider the particular choice $a = (0,0,1)$.}\label{fig}
\end{figure}

In~\cite{ghirardi3}, as a measure of the maximal departure from quantum expectations that crypto-nonlocal
hidden variables models models can provide, we have used the variance of the variable $f_{\psi}
(a, \tau)$ over the distribution $\rho_{\psi}(\tau)$:
\begin{equation}\label{distance}
    \delta_{\psi} (a) = \int \Big( f_{\psi} (a, \tau) - \langle x (a) \rangle_{\psi} \Big)^2 \, \rho_{\psi}(\tau) d \tau.
\end{equation}
When the system consist of a pair of qubits, we have expressed the upper bound for this quantity for generic models as
\begin{equation}\label{constraint3}
    \delta_{\psi} (a) \leq \cos{\theta} - \langle x (a) \rangle_{\psi}^2.
\end{equation}
In Fig.~\ref{fig} we show the dependence of this constraint by entanglement, and plot the
corresponding curve for the specific model described in~\cite{ghirardi2}.

\section{Final remarks and conclusions}

In this contribution we have criticized the form of the free
will assumption which has been recently adopted to derive some
striking results, in particular that (i) quantum mechanics is
a complete theory, and (ii) the quantum state vector $\psi$ is necessarily
in one-to-one correspondence with the elements of reality of the theory.
To strengthen our argument, we have provided a different definition of
free will, which makes clear that the former free will assumption
actually encodes more than the observers' free choice (that is, the
condition that the theory is non-signalling, and the staticity of the
information supplementing $\psi$). We believe that both approaches
to free will do not represent the correct necessary and sufficient
condition for this assumption, and we conjecture that this condition cannot
be simply expressed through simple expressions involving conditional
probabilities.

Therefore, ontological models which are compatible but possibly experimentally
distinguishable from quantum mechanics are possible. Since these models should
necessarily be non-signalling, we have suggested that they are given by the class
of crypto-nonlocal hidden variables theories, and we have provided a simple example.
Finally, we have described an upper bound for the local averages of any deterministic
ontological theory for quantum mechanics in the case a pair of qubits. These results
suggest that crypto-nonlocal hidden variables theories represent a relevant tool in the
study of non-locality.

\ack This research is partially supported by the ARO MURI grant W911NF-11-1-0268.


\section*{References}

\begin{thebibliography}{99}


\bibitem{harrigan}
N. Harrigan and R.W. Spekkens, Found. Phys. 40, 125 (2010)


%\bibitem{bell}
%J.S. Bell, "Beables for quantum field theory," CERN-TH
%4035/84, Aug. 2, 1984;
%
%\bibitem{shimony}
%A. Shimony, M.A. Horne and J.F. Clauser, "Comment on 'The
%Theory of Local Beables'," from Epistemological Letters
%Hidden Variables and quantum Uncertainty, F. Bonsack,
%ed., No. 9 (1976)

\bibitem{colbeck2}
R. Colbeck and R. Renner, Nature Comm. 2, 411 (2011)

\bibitem{colbeck}
R. Colbeck and R. Renner, Phys. Rev. Lett. 108, 150402 (2012)

\bibitem{pusey}
M.F. Pusey, J. Barrett and T. Rudolph, Nature Phys. 8,
476 (2012)

\bibitem{lewis}
P.G. Lewis, D. Jennings, J. Barrett and T. Rudolph, Phys. Rev. Lett. 109, 150404 (2012)

\bibitem{ghirardi}
G.C. Ghirardi and R. Romano, Foundations of Physics: Volume 43, Issue 7 881 (2013)

\bibitem{bell}
J.S. Bell, {\it Free variables and local causality}, in {\it Speakable
and unspeakable in quantum mechanics}, Cambridge
University Press (1987)

\bibitem{leggett}
A.J. Leggett, Found. Phys. 33, 1469 (2003)


%\bibitem{bell}
%J.S. Bell, Rev. Mod. Phys. 38, 447 (1966)
%
%\bibitem{kochen}
%S. Kochen and E.P. Specker, J. of Math. and Mech. 17, 59 (1967)
%
%\bibitem{pusey}
%M.F. Pusey, J. Barrett and T. Rudolph, Nature Phys. 8, 476 (2012)
%
%\bibitem{colbeck2}
%R. Colbeck and R. Renner, Phys. Rev. Lett. 108, 150402 (2012)
%
%\bibitem{lewis}
%P.G. Lewis, D. Jennings, J. Barrett and T. Rudolph, Phys. Rev. Lett. 109, 150404 (2012)
%
%\bibitem{leggett}
%A.J. Leggett, Found. Phys. 33, 1469 (2003)
%
%\bibitem{colbeck3}
%R. Colbeck and R. Renner, Phys. Rev. Lett. 101, 050403 (2008)
%
%\bibitem{branciard}
%C. Branciard, A. Ling, N. Gisin, C. Kurtsiefer, A. Lamas-Linares and V. Scarani, Phys. Rev. Lett. 99, 210407 (2007)
%
%\bibitem{ghirardi}
%G.C. Ghirardi and R. Romano, Phys. Rev. A 86, 022107 (2012)
%
%\bibitem{brans}
%C. Brans, Int. J. Theor. Phys. 27, 219 (1988)
%
%\bibitem{hall}
%M.J.W. Hall, Phys. Rev. Lett. 105, 250404 (2010)
%
%\bibitem{hardy}
%L. Hardy, Stud. Hist. Phil. Mod. Phys. 35, 267 (2004)
%
%\bibitem{montina}
%A. Montina, Phys. Rev. A 83, 032107 (2011)
%
%\bibitem{montina2}
%A. Montina, Phys. Lett. A 375, 1385 (2011)
%
%\bibitem{ghirardi2}
%G.C. Ghirardi and R. Romano, {\it About possible extensions of quantum theory}, in preparation
%
\bibitem{ghirardi2}
G.C. Ghirardi and R. Romano, Phys. Rev. A 85, 042102 (2012)

\bibitem{ghirardi3}
G.C. Ghirardi and R. Romano, Phys. Rev. Lett. 110, 170404 (2013)

%\bibitem{niel} M.A. Nielsen and I.L. Chuang, {\it Quantum Computation and Quantum
%Information}, Cambridge, 2000
%
%\bibitem{shor} P. W. Shor, Phys. Rev. A 52, R2493 (1995)
%
%\bibitem{knil} E. Knill and R. Laflamme, Phys. Rev. A 55, 900 (1997)
%
%\bibitem{benn} C.H. Bennett, D.P. DiVincenzo, J.A. Smolin, and W.K. Wootters,
%Phys. Rev. A 54, 3824 (1996)
%
%\bibitem{benn2} C.H. Bennett, G. Brassard, C. Crepeau, R. Jozsa, A. Peres and W.K. Wootters,
%Phys. Rev. Lett. 70, 1895 (1993)
%
%\bibitem{bosc} D. Boschi, S. Branca, F. De Martini, L. Hardy, and S. Popescu,
%Phys. Rev. Lett. 80, 1121 (1998)
%
%\bibitem{bowe} G. Bowen and S. Bose, Phys. Rev. Lett. 87, 267901 (2001)
%
%\bibitem{benn3} C. H. Bennett, G. Brassard, S. Popescu, B. Schumacher, J. A. Smolin,
%and W. K. Wootters, Phys. Rev. Lett. 76, 722 (1996)
%
%\bibitem{mor} T. Mor, P. Horodecki, e-print quant-ph/9906039
%
%\bibitem{son} W. Son, J. Lee, M.S. Kim, and Y.-J. Park, Phys. Rev. A 64, 064304 (2001)
%
%\bibitem{modl} J. Modlawska and A. Grudka, Phys. Rev. Lett. 100, 110503 (2008)
%
%\bibitem{brun} T. Brun, I. Devetak, and M.H. Hsieh, Science 314, 436 (2006)
%
%\bibitem{jami} A. Jamiolkowski, Rep. Math. Phys. 3, 275 (1972)
%
%\bibitem{roma} R. Romano and P. Van Loock, {\it Quantum control of noisy channels}, in preparation
%
%
%
%








\end{thebibliography}
\end{document}